\pdfoutput=1
\documentclass[12pt]{iopart}
\usepackage{color}
\usepackage{verbatim}
\usepackage{graphicx}
\usepackage{caption}
\usepackage{cleveref}
\usepackage{cite}
\usepackage{xr}
\usepackage{textcomp}

\newcommand{\phys}{\emph{Physarum} }

\usepackage{amssymb,latexsym}
\usepackage{url}

\definecolor{verde}{rgb}{0,0.5,0}

\newcommand{\mylab}[3]{\raisebox{#2}[0mm][0mm]{%
\makebox[0mm][l]{\hspace*{#1}\textbf{#3}}}}
\def\spacce#1{\hskip #1pt}
\def\drawline#1#2{\raise 2.5pt\vbox{\hrule width #1pt height #2pt}}

\def\circle{$\circ$\nobreak }
\def\linecir{\hbox%
{\drawline{8}{.5}\spacce{2}\circle\spacce{2}\drawline{8}{.5}}\nobreak}
\def\trian{\raise 1.25pt\hbox{$\scriptstyle\triangle$}\nobreak}
\def\leftrian{\raise 0.75pt\hbox{$\displaystyle\triangleleft$}\nobreak}
\def\rightrian{\raise 0.75pt\hbox{$\displaystyle\triangleright$}\nobreak}
\def\linetri{\hbox%
{\drawline{8}{.5}\spacce{2}\trian\drawline{8}{.5}}\nobreak}
\def\dtrian{\raise 1.25pt\hbox%
{$\scriptscriptstyle\bigtriangledown$}\nobreak}

\def\squar{\raise 1.25pt\hbox{$\scriptstyle\Box$}\nobreak}
\def\linesquar{\hbox%
{\drawline{8}{.5}\spacce{2}\squar\drawline{8}{.5}}\nobreak}
\def\diamon{\raise 1.25pt\hbox{$\scriptstyle\diamond$}\nobreak}

\def\solidsquar{$\blacksquare$\nobreak}

\def\solidcircle{$\bullet$\nobreak}

\def\dd{{\, \rm{d}}}

\def\beq{\begin{equation}}
\def\eeq{\end{equation}}

\def\aaa{{\it a}}
\def\bbb{{\it b}}
\def\ccc{{\it c}}
\def\ddd{{\it d}}
\def\eee{{\it e}}
\def\fff{{\it f}}
\def\ggg{{\it g}}
\def\hhh{{\it h}}

\def\citalajim03{Del \'Alamo \& Jim\'enez (2003)}
\newcommand\ie{{\it i.e.}\;}
\newcommand\eg{{\it e.g.}\;}

\begin{document}

\title[Self-organized dynamics of migrating \phys fragments]
{Self-organized mechano-chemical dynamics 
in amoeboid locomotion of \phys fragments}
\author{Shun Zhang$^1$, Robert D.\ Guy$^2$, Juan C.\ Lasheras$^{1,3,4}$ \\ and Juan C.\ del \'Alamo$^{1,4}$}
\address{$^1$Mechanical and Aerospace Engineering Department, University of California San Diego
\newline $^2$Department of Mathematics, University of California Davis
\newline $^3$Department of Bioengineering, University of California San Diego
\newline $^4$Institute for Engineering in Medicine, University of California San Diego}
\ead{shz019@ucsd.edu, jalamo@ucsd.edu}
\vspace{10pt}
\begin{indented}
\item[]March 2017
\end{indented}

\begin{abstract}
The aim of this work is to quantify the spatio-temporal dynamics of flow-driven
amoeboid locomotion in small ($\sim$100 $\mu$m) fragments of the true slime
mold \phys {\it polycephalum}.  In this model organism, cellular contraction
drives intracellular flows, and these flows transport the chemical signals that
regulate contraction in the first place.  As a consequence of these non-linear
interactions, a diversity of migratory behaviors can be observed in migrating
\phys fragments.  To study these dynamics, we measure the spatio-temporal
distributions of the velocities of the endoplasm and ectoplasm of each
migrating fragment, the traction stresses it generates on the substratum, and
the concentration of free intracellular calcium.  Using these unprecedented
experimental data, we classify migrating \phys fragments according to their
dynamics, finding that they often exhibit spontaneously coordinated waves of
flow, contractility and chemical signaling.  We show that \phys fragments
exhibiting symmetric spatio-temporal patterns of endoplasmic flow migrate
significantly slower than fragments with asymmetric patterns.  
In addition, our joint measurements of ectoplasm velocity and traction stress
at the substratum suggest that forward motion of the ectoplasm is enabled by a
succession of stick-slip transitions, which we conjecture are also organized in
the form of waves.
Combining our experiments with a simplified convection-diffusion model, we show
that the convective transport of calcium ions may be key for establishing and
maintaining the spatio-temporal patterns of calcium concentration that regulate
the generation of contractile forces.  
%
\end{abstract}

\maketitle

\section{Introduction}
\label{intro}

Amoeboid locomotion is a fast type of cellular locomotion that involves large
shape changes mediated by cell contractility, and does not require
biochemically regulated adhesion to the extracellular environment
\cite{Charras:2008gf,Lammermann:2008ib,yoshida2006dissection,alvarez:2014}.
In addition to its conspicuous biomedical applications, the study of amoeboid
locomotion has been recently applied to the bio-mimetic design of fluid filled,
highly deformable robots \cite{Piovanelli:2012el,umedachi2013true}.
%
%
Amoeboid organisms such as {\it Amoeba proteus} and {\it Physarum polycephalum}
are particularly interesting model organisms for biomimetic design because they
develop significant intracellular pressure-driven flows
\cite{kamiya1959,allen:1978cytoplasmic}.
Because diffusion across these giant cells is slow, intracellular flows are
important not only to drive motility but also for the transport of chemical
signals and nutrients (see \S \ref{sec:ca2plus} below and
\cite{goldstein:2015physical}).  The non-linear feedback between
pressure-driven flow, molecular transport and cell contractility can lead to
rich dynamics that differ from those observed in smaller cells, and which are
yet poorly understood.  

This study examines the spatio-temporal dynamics of flow driven amoeboid
locomotion in the true slime mold \phys {\it polycephalum}.
The \phys plasmodium is a multi-nucleated slime mold that is composed
of a gel-like ectoplasm and a sol-like endoplasm \cite{aldrich2012cell}.
During locomotion, the endoplasm flows back and forth in a periodic manner,
which is customarily characterized as ‘shuttle flow’
\cite{kamiya1959,Matsumoto:2008jf}.  This flow is driven by periodic
contractions of cross linked actomyosin fibrils in the ectoplasm
\cite{nagai1978cyclic,Lewis20141359,Rieu20150099}.  The contraction is
regulated by waves of calcium ions
\cite{ridgway1976oscillations,yoshimoto1981simultaneous}, whose propagation is
notably influenced by the endoplasmic flow
\cite{yoshiyama2010calcium,zhang:2016coordinations}. 
The interactions between these physical phenomena can be associated with the
complex spatio-temporal patterns observed in a variety of \phys preparations,
including non-locomoting protoplasmic droplets, and locomoting plasmodial
fragments of small ($\sim 100\, \mu$m) \cite{Lewis20141359} and intermediate
($\sim 1\, \mu$m) size \cite{rodiek2015patterns}.
These experiments reported homologous spatio-temporal patterns for a range of
different geometries, biological strains and environmental conditions, implying
that the spatio-temporal coordination of motility in \phys plasmodia could be
achieved via remarkably robust physical mechanisms.  However, investigating the
details of these mechanisms has been difficult since the vast majority of
previous experiments recorded a limited amount of data, namely the fragment
thickness as estimated indirectly from image brightness.  In particular, there
is very limited information about the spatio-temporal dynamics of calcium ions
and their relation with endoplasmic flow and ectoplasmic contractility
\cite{zhang:2016coordinations}.

Mathematical models facilitate investigating the spatio-temporal coordination
of \phys motility by integrating the available experimental data into
quantitative frameworks including variables that may be hard to measure
experimentally.
Various models have been constructed under this premise, and the numerical
results have reproduced a variety of experimentally observed spatio-temporal
patterns \cite{Radszuweit:2013ku,Radszuweit:2014be,Lewis20141359,Alonso:2016}.
However, these models have been so far limited to fixed simple geometries or
have neglected key aspects of the mechano-chemical feedback present in
locomoting \phys fragments.

This paper presents novel multi-channel measurements of mechanical and chemical
variables in plasmodial fragments of \phys {\it polycephalum} undergoing
directional migration.
These measurements provide simultaneous spatio-temporal maps of the contractile
forces generated by the fragments on their substratum, the velocities of their
endoplasm and ectoplasm, and the distribution of free intracellular calcium
concentration ([Ca$^{2+}$]i). 
The experimental data are analyzed to study how a biological system like \phys
coordinates the generation of mechanical forces with their shape changes and
internal flows via adhesion to its substratum, and how these pressure driven
flows transport the chemical signals that regulate force generation in the
first place.  The ultimate goal of the analysis is to understand how these
phenomena are spontaneously organized to enable the directional flow driven
locomotion of amoeboid organisms.
%

\section{Methods}
\label{sec:methods}

\subsection{Preparation of \phys fragments}
Motile \phys fragments of approximately $500$ \textmu m in length were prepared
as in our previous study \cite{Lewis20141359}.  {\it Physarum} plasmodia were
grown on 1\% agar gel (Granulated; BD) using $150 \times 15$ mm culture plates
(BD), fed with oat flakes (QUAKER) and kept in a dark humid environment at room
temperature.  Small portions of $\sim 0.2 \times 0.2$ mm$^2$ were cut from the
parent plasmodia, transfered to collogen coated polyacrylamide (PA) gels
embedded with fluorescent beads.  A cap made of agarose was placed over the
\phys fragments immediately after.  After several hours, the fragments adapted
to tadpole like shape and perfomed directed migration, with noticeable
intracellular streaming.

\subsection{Gel Fabrication}
Collagen-coated PA gels were prepared for traction force microscopy as
previously described \cite{DelAlamo:2013di}. The gel was $ \sim 1.5$ mm thick
and consisted of two layers, the top layer was thin ($\sim 10$ \textmu m) and
contained $0.5$ \textmu m florescent beads (FluoSperes; Molecular Probes).
The gels were fabricated using 5\% acrylamide and 0.3\% bisacrylamide (Fisher
BioReagents), resulting in a Young's modulus value equals to 8.73 kPa
\cite{Tse:2010ft}.  The Poisson's ratio of the gel was measured to be 0.46
using the forces generated by the \phys fragments themselves, following an
elastographic  traction force microscopy method recently developed by our group
\cite{AlvarezGonzalez:2017}.
The cap, made of 0.8\% agarose with thickness of $3$ mm, prevented the PA gel
from drying out and generate gentle confinement (a $\sim 30$ Pa and a $\sim 1$
KPa Young's modulus) to facilitate the measurement of intracellular flow.
More details about the effect of the agar cap on our measurements and the
migration of \phys fragments can be found in
\cite{Lewis20141359,AlvarezGonzalez:2017}.

\subsection{Microscopy}
A Leica DMI 6000B inverted microscope and a PC running Micro-Manager software
were used for image acquisition \cite{Edelstein:2010gf}.  Time-lapse sequences
were acquired at 16X in both bright-field and fluorescent-field.  First, 10
images were acquired in the bright field at a frame rate of 5 Hz for flow
quantification. Then, a 40-image fluorescence z-stack ($\Delta z = 1$ \textmu
m) was acquired over 10 sec for traction force microscopy.  This $12$-second
acquisition cycle was repeated until the cell moved out of the field of view,
providing quasi-simultaneous recordings of intracellular streaming and traction
stresses, given that the variables oscillate with a much longer period of $\sim
100$ sec \cite{Matsumoto:2008jf}.
%

\subsection{Flow Quantification}
The cytoplasm of \phys amoebae is densely packed with intracellular vesicles,
which were used as fiduciary markers to quantify the intracellular streaming
velocity by particle image velocimetry (PIV)
\cite{Matsumoto:2008jf,Willert:1991hq}.  The intracellular domain can be
separated as endoplasmic flow phase and ectoplasmic gel phase with respective
characteristic velocities of $10$ \textmu m/s and $0.15$ \textmu m/s.
Different algorithms were used to determine the velocities range across 2
orders of magnitude.  For $\vec{V}_{sol}$, we pre-process the raw image
sequences using high-pass, band-pass and low-pass temporal filters as described
in our previous paper \cite{Lewis20141359}.
Then we ran an in-house PIV algorithm on each filtered image sequences and
asigned the velocity vector resulting from the sequences that maximizes the PIV
signal-to-noise ratio at each point.  As for $\vec{V}_{gel}$, we ran PIV on
image pair consists of the first and last image in bright field of each
acquisition cycle.  The rather long time interval (1.8 second) allowed us to
detect the low velocities of the ectoplasmic gel phase.  Points with velocity lower than
$0.2$ \textmu m/s were considered as ectoplasm.  The PIV interrogation window size
and spacing were respectively 32 and 8 pixels, yielding a spatial resolution of
$6.5$ \textmu m.

\subsection{Traction force microscopy}
The $3D$ deformation of the PA substrate was measured at its top surface on
which the \phys amoebae were migrating as reported by del \'Alamo et al.\
\cite{DelAlamo:2013di}.  Each instantaneous fluorescence $z$-stack was
cross-correlated with a reference $z$-stack which was recorded at the end of
experiment once the amoebae moved out of the field of view.  Using these
measurements as boundary conditions, we computed the three-dimensional
deformation field in the whole polyacrylamide substrate by solving the
elasto-static equation.  We then compute the traction stress $\vec \tau =
(\tau_{xz},\tau_{yz},\tau_{zz})$ exerted by the cell on the substrate using
Fourier TFM methods described elsewhere \cite{DelAlamo:2007jm,DelAlamo:2013di}.
The spatial resolution of $\vec \tau$ was $13$ \textmu m in x,y and $1$ \textmu m in z.

\subsection{Measurement of free calcium concentration}
Single-wave length calcium indicators like Calcium Green-1 (Molecular Probes)
exhibit an increase in fluorescence upon binding calcium ions and have been
successfully applied to monitor the dynamics of [Ca$^{2+}$]i.
However, the recorded intensity of these
indicators can vary with other factors such as cell thickness.  In our
experiments, the local thickness of \phys fragments can vary up to $50 \%$
during migration.  Dual-wavelength ratiometric dyes, with distinct spectra of
calcium free and calcium bond forms, can be used to minimize the effect of
variation in cell thickness.  However, dual-wavelength dyes require excitation
in the UV range, for which \phys fragments exhibit significant
auto-fluorescence.  To solve this problem, Texas Red (Molecular Probes), which
is a calcium insensitive fluorescent dye, is co-injected into the sample
together with Calcium Green-1.  The ratio of fluorescent intensity between
calcium sensitive dye and background dye are used to monitor the variation of
[Ca$^{2+}$]i.  Both dyes were dextran-conjugated and had a molecular weight of
10 kDa, which dramatically reduced leakage and compartmentalization compared to
their non-conjugated forms.  The dyes were coinjected into the parent mold
under a Nikon SMZ-10 microscope using a PM 1000 cell micro-injection system
(MicroData Instrument, Inc).  Time-lapse sequences were acquired under 20X in
bright-field, FITC and TRITC.  10 images in bright field were acquired first at
5 Hz for flow quantification, followed by one snapshot in FITC for Calcium
Green-1 and another one in TRITC for Texas Red.  This 5-second acquisition
cycle was repeated for at least 10 minutes, allowing us to obtain a
quasi-simultaneous quantification of intracellullar flow and [Ca$^{2+}$]i
during \phys migration.  Preliminary results obtained from this acquisition
protocol were presented in \cite{zhang:2016coordinations}.

\subsection{Fragment shape statistics}
%
Cell contours are extracted from bright-field microscopy time-lapse sequences
as described previously \cite{DelAlamo:2007jm}.  Raw images are digitally
thresholded, eroded and dilated in order to obtain a time-dependent scalar
field $\Omega_c(t,x,y)$ containing ones inside the fragment and zeroes outside
of it.  
The statistical distributions of fragment shape are determined from
$\Omega_c(t,x,y)$ following the method outlined in ref. \cite{meili:2010}.  At
each instant of time, $\Omega_c(t,x,y)$ is rotated so that the major axis of
the fragment coincides with the $x$ direction, and translated so that the
origin $(x,y)=(0,0)$ is set at the centroid of the fragment. 
The probability density function of a point belonging to the interior of the
fragment in this reference frame is calculated simply as 
\begin{equation}
%
P(x,y)=\frac 1 {N_c} \sum_{i=1}^{N_c} \frac 1 {N_{t,i}} \sum_{j=1}^{N_{t,i}}
\Omega_{c,i}(t_j,x,y),
%
\end{equation}
where $N_c$ is the number of cells and $N_{t,i}$ is the number of temporal
observations corresponding to the $i$-th cell.  According to this definition,
the median cell shape is defined by the iso-contour $P(x,y)=0.5$.

\subsection{Kymographic representation}
To facilitate the analysis of the spatio-temporal organization of endoplasmic
flow, traction stresses, free intracellular calcium, etc., we generated
kymographs for the quantities of interest.  
We followed the approach introduced by Bastounis \etal \cite{Bastounis:2014gg}
for migrating amoeboid cells.
At each instant of time ($t$) the major axis of cell is aligned vertically
($x$), and the measured quantity ($\mathbf{q}(t,x,y)$) is projected and the
averaged over the cross section of the \phys fragment, \ie
\begin{equation}
%
\overline q(t,x) = 
\frac{\int \Omega_c(t,x,y) \mathbf{q}(t,x,y) \cdot \mathbf{u_x} \,\dd y}{\int
\Omega_c(t,x,y) \, \dd y},
\label{eq:kymo_def}
%
\end{equation} 
where $\Omega_c$ denotes the interior of the \phys fragment, and $\mathbf{u_x}$
is a unit vector oriented towards the fragment's front.  
Plotting two-dimensional maps of $\overline q$ produces kymographs that can
reveal patterns of organization in the spatio-temporal dynamics of migrating
\phys fragments.


\section{Results and Discussion}
\subsection{The spatio-temporal organization of endoplasmic flow and traction
stresses reveals distinct dynamical modes in migrating \phys fragments}
A few hours after seeding the \phys fragments on PA gel, we observed that
fragments of diameter larger than $\sim 100$ \textmu m  performed persistent
locomotion.  Fragments larger than $\sim 500$ \textmu m formed complex branched
structures markedly different from an amoeboid shape, and were not considered
in this study.  We focused our investigation on fragments of size $\sim 300$
\textmu m, which generally adopted a tadpole-like shape, with a more rounded
head and a tapering tail.
\begin{figure}[tb!]
\begin{center}
\vspace{2ex}
\mylab{0.00\textwidth}{0.295\textwidth}{{\footnotesize (\aaa)}}
\mylab{0.07\textwidth}{0.295\textwidth}{{\scriptsize t=$1070$s}}
\mylab{0.2\textwidth}{0.295\textwidth}{{\scriptsize t=$1081$s}}
\mylab{0.34\textwidth}{0.295\textwidth}{{\scriptsize t=$1092$s}}
\mylab{0.48\textwidth}{0.295\textwidth}{{\scriptsize t=$1103$s}}
\mylab{0.62\textwidth}{0.295\textwidth}{{\scriptsize t=$1114$s}}
\mylab{0.77\textwidth}{0.295\textwidth}{{\scriptsize t=$1125$s}}
\includegraphics[width=0.9\textwidth]{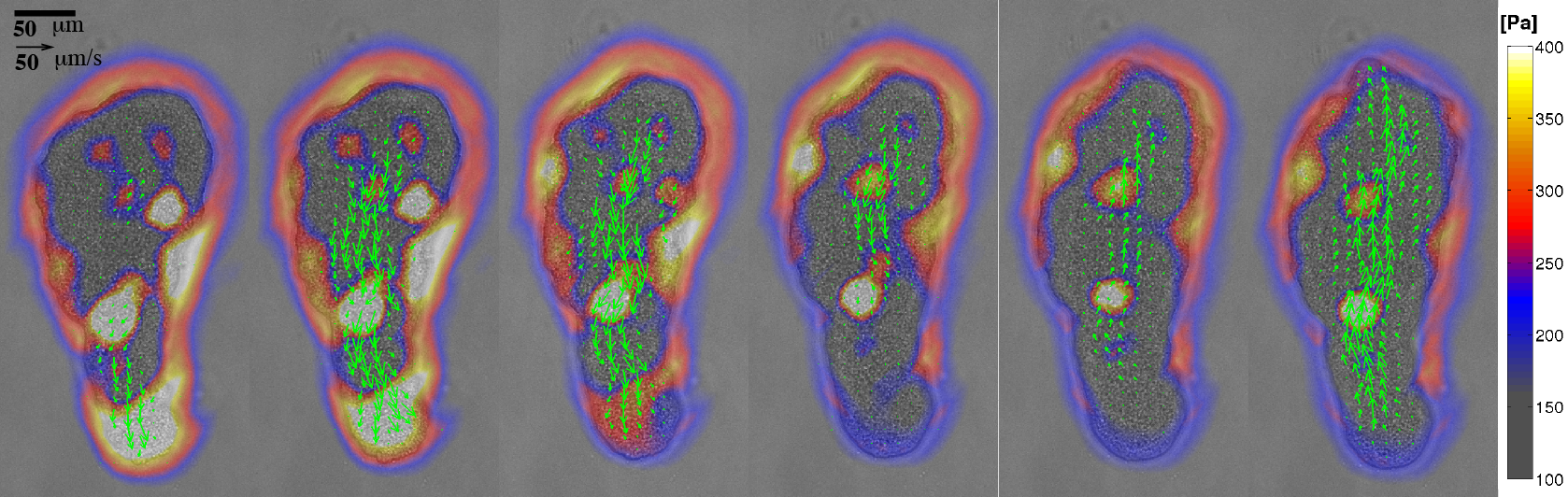}
\\[4ex]
\mylab{0.02\textwidth}{0.21\textwidth}{{\footnotesize (\bbb)}}
\mylab{0.06\textwidth}{0.21\textwidth}{{\tiny t=$396$s}}
\mylab{0.16\textwidth}{0.21\textwidth}{{\tiny t=$416$s}}
\mylab{0.26\textwidth}{0.21\textwidth}{{\tiny t=$436$s}}
\mylab{0.36\textwidth}{0.21\textwidth}{{\tiny t=$455$s}}
\includegraphics[height=0.2\textwidth]{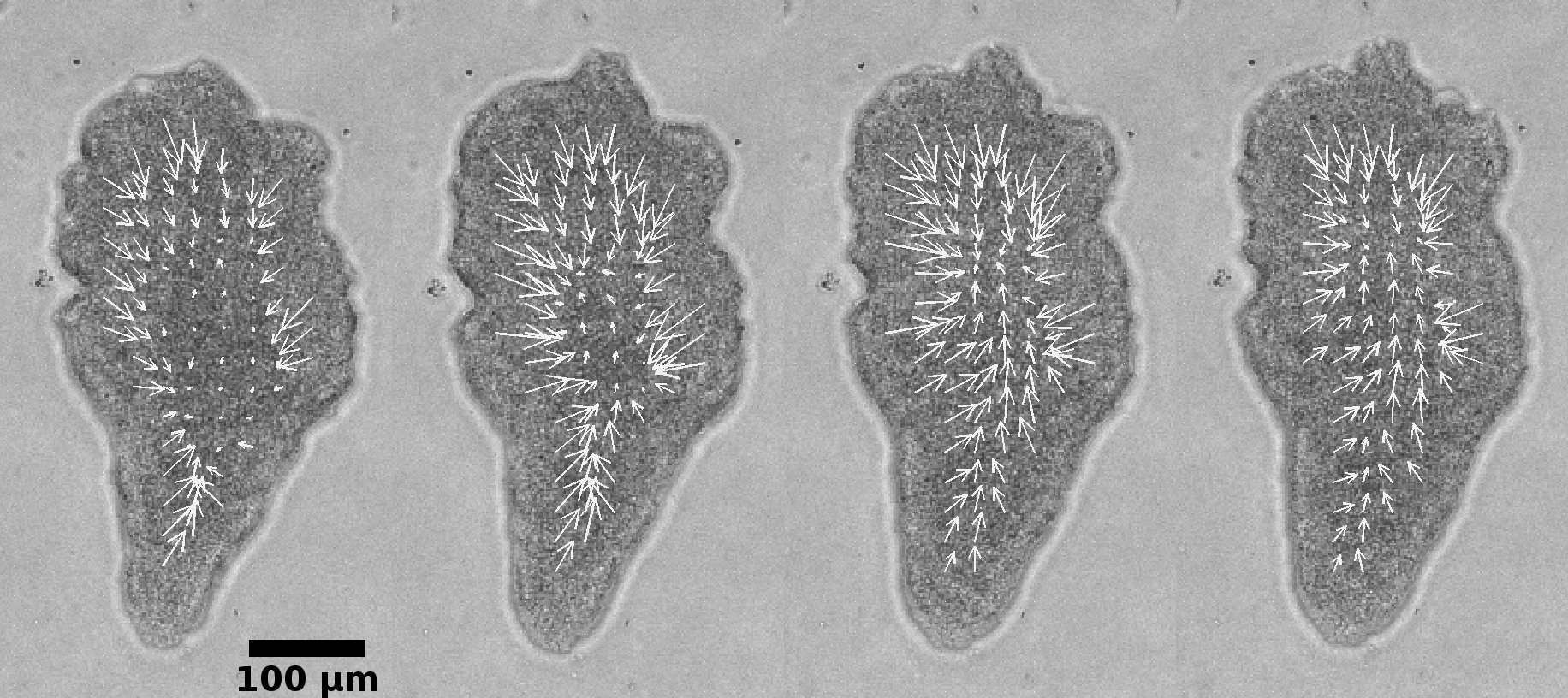}
\mylab{0.02\textwidth}{0.21\textwidth}{{\footnotesize (\ccc)}}
\mylab{0.06\textwidth}{0.21\textwidth}{{\tiny t=$462$s}}
\mylab{0.16\textwidth}{0.21\textwidth}{{\tiny t=$482$s}}
\mylab{0.26\textwidth}{0.21\textwidth}{{\tiny t=$507$s}}
\mylab{0.36\textwidth}{0.21\textwidth}{{\tiny t=$527$s}}
\includegraphics[height=0.2\textwidth]{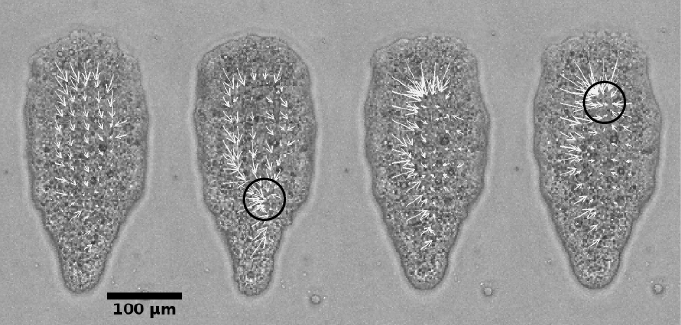}
\caption{(\aaa) Instantaneous endoplasmic flow speed and traction stresses
exerted on the substrate by a migrating \phys fragment.  Arrows exhibit the
flow speed and colormap shows the magnitude of traction stresses. (\bbb, \ccc)
Instantaneous traction stresses exerted by two \phys fragments with different
dynamical behaviors, with the stress vectors along the cell boundary removed.
Black circles indicate the location of contraction centers.  } 
\label{fig:snapshots0}
\end{center}
\end{figure}

Directional locomotion of \phys fragments requires the spatio-temporal
coordination of endoplasmic flows and traction stresses exerted on a contact
surface \cite{Lewis20141359}.  Most of the fragments analyzed in this study
developed organized endoplasmic flows that oscillated between forward and
backward motion with a well defined periodicity (Figure
\ref{fig:snapshots0}\aaa).  The traction stresses exerted by these locomoting
fragments oscillated with a similar period, all the time showing an inward
contractile pattern with larger stresses along the cell periphery.  
This pattern has been proposed to be analogous to a surface tension
\cite{Rieu20150099,delanoe20104d}, and has been recently linked to the cortical
F-actin filaments and their cross-linkers in {\it Dictyostelium} amoebae
\cite{AlvarezGonzalez:2015km}.
Removing the stress vectors near each fragment's boundary renders the traction
stress generated under the fragment's body easier to analyze, and unmasks waves
of contraction with distinct spatio-temporal dynamics
(\Cref{fig:snapshots0}\bbb, \ccc).

\begin{figure}[tb!]
\includegraphics[width=0.95\textwidth]{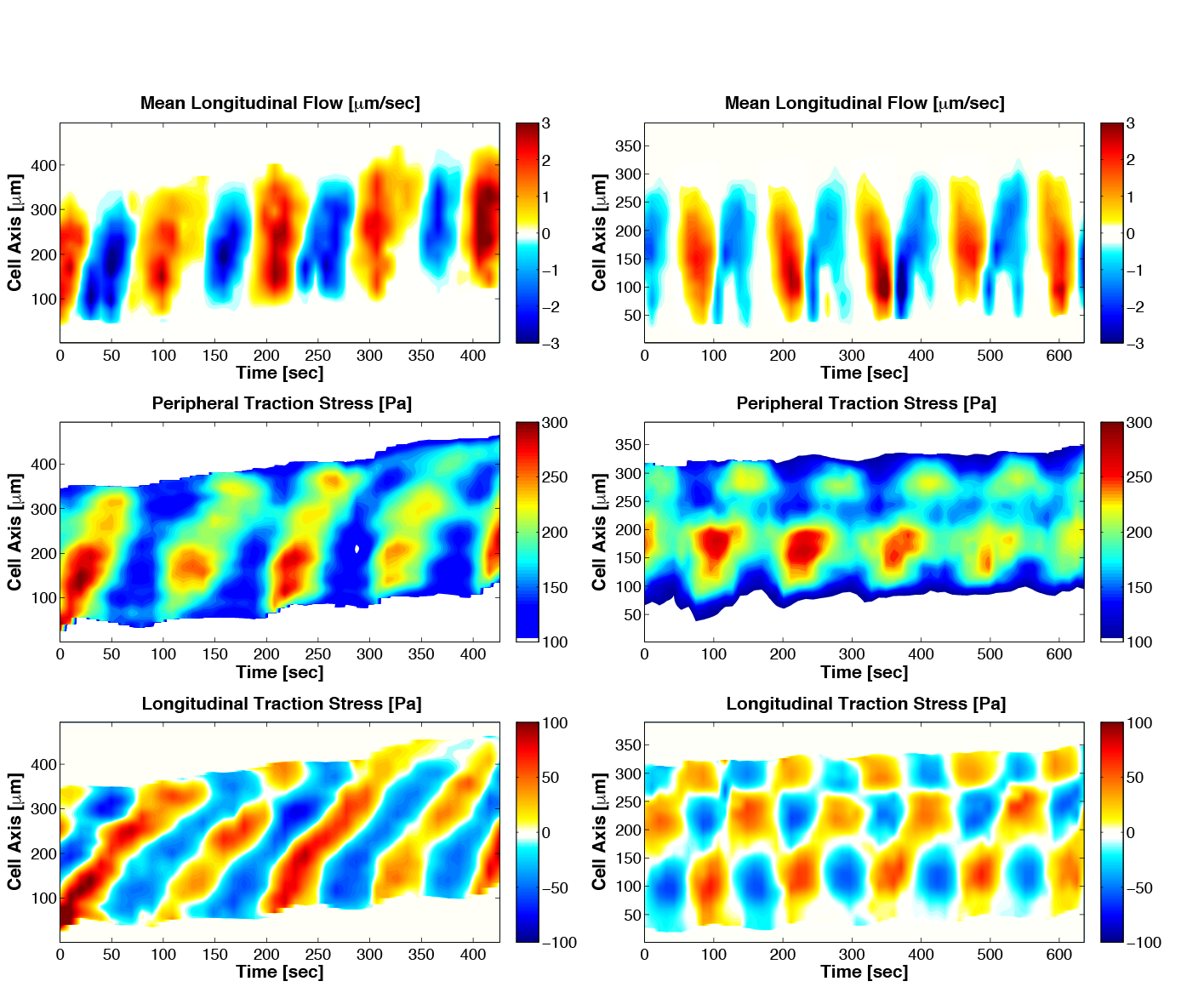}
\mylab{-0.9\textwidth}{0.68\textwidth}{{\footnotesize (\aaa) }}
\mylab{-0.9085\textwidth}{0.438\textwidth}{{\footnotesize (\bbb) }}
\mylab{-0.915\textwidth}{0.193\textwidth}{{\footnotesize (\ccc) }}
\mylab{-0.445\textwidth}{0.68\textwidth}{{\footnotesize (\ddd) }}
\mylab{-0.455\textwidth}{0.438\textwidth}{{\footnotesize (\eee) }}
\mylab{-0.46\textwidth}{0.193\textwidth}{{\footnotesize (\fff) }}
\caption{Kymographs of longitudinal endoplasm flow velocity (\aaa, \ddd), 
  peripheral traction stress (\bbb, \eee), and longitudinal traction stress 
  (\ccc, \fff) for a peristaltic (\aaa–-\ccc) and 
  an amphistaltic (\ddd–-\fff) \phys fragment.}
\label{fig:kymo1}
%
\end{figure}

The most common organized pattern consisted of traveling waves that propagated
forward along the center line of the motile fragment (Figure
\ref{fig:kymo1}\aaa--\ccc).  We labeled this migration mode as {\it
peristaltic} because their motion was driven by forward traveling waves of
contraction and relaxation (Figures \ref{fig:snapshots0}\bbb\; and
\ref{fig:kymo1}\bbb,\ccc).  
This terminology is based on previous studies of \phys migration
\cite{Matsumoto:2008jf,aldrich2012cell,alim2013random}, in which peristaltic
fragments were designated by analyzing the dynamics of fragment width change
rather than their force generation dynamics.
In \phys fragments undergoing peristaltic locomotion, both the forward and
backward endoplasmic flow waves are generated from the tail and propagate
forward in an approximately linear fashion.  This mode has drawn more attention
in previous studies because it occurs more often and leads to faster migration
than other modes
\cite{Matsumoto:2008jf,Lewis20141359,romanovskii1995physical,kamiya1959}.
However, it is not the only migration mode of \phys locomotion with organized
spatio-temporal dynamics.

We also observed a less frequent yet distinct mode of locomotion in which the
head and tail contracted and relaxed in an anti-phase manner, and which we
named the {\it amphistaltic} mode.   This mode sustains waves of forward and
backward endoplasmic flow that alternate periodically, similar to the
peristaltic mode.  However, in \phys fragments undergoing amphistaltic
locomotion, the waves of forward endoplasmic flow originate at the fragment's
front and propagates backward, whereas the waves of backward flow originate at
the fragment's back and travel forward.  This dynamics leads to evident
`V'-shaped patterns in the flow kymograph (Figure \ref{fig:kymo1}\ddd).
The instantaneous spatial patterns of traction stresses in amphistaltic \phys
fragments showed inward contraction similar to peristaltic ones (Figure
\ref{fig:snapshots0}\ccc).  However, the traction stress kymographs revealed
remarkable differences in their spatio-temporal dynamics.  Instead of traveling
waves, amphistaltic fragments sustained standing waves of traction stress with
alternating peaks and valleys at the front and rear of the fragment (Figure
\ref{fig:kymo1}\eee, \fff).  Consistently, traction stress snapshots of
amphistaltic fragments show localized contraction centers in the front and rear
part of the fragment (black circles in Figure \ref{fig:snapshots0}\ccc).
This pattern of contraction resembles that of the \phys ``dumbbells''
previously described by others
\cite{Gawlitta:1980,Brix:1987ta,Kukulies:1987ed,Rieu20150099}.  These dumbbells
form two thick round heads connected by a tube that contract alternatively
while the fragment stays in place.  We also observed a few contractile
dumbbells in our experiments.  However, the amphistaltic \phys fragments
reported here always adopted a tadpole-like shape and were able to move
persistently.
%

\begin{figure}[tb!]
\includegraphics[width=0.95\textwidth]{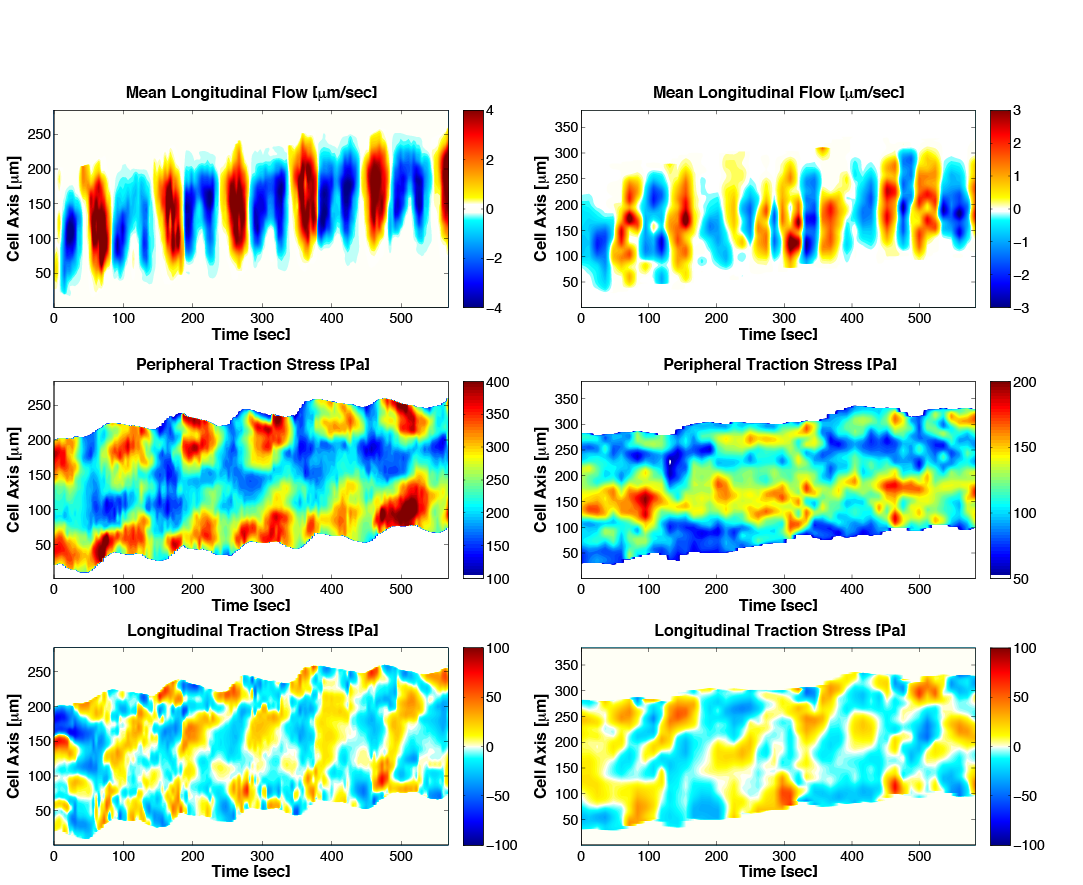}
\mylab{-0.9\textwidth}{0.67\textwidth}{{\footnotesize (\aaa) }}
\mylab{-0.9085\textwidth}{0.43\textwidth}{{\footnotesize (\bbb) }}
\mylab{-0.915\textwidth}{0.19\textwidth}{{\footnotesize (\ccc) }}
\mylab{-0.445\textwidth}{0.67\textwidth}{{\footnotesize (\ddd) }}
\mylab{-0.455\textwidth}{0.43\textwidth}{{\footnotesize (\eee) }}
\mylab{-0.46\textwidth}{0.19\textwidth}{{\footnotesize (\fff) }}
%
%
\caption{Kymographs of flow and traction stress of two \phys that exhibited
uncommon spatio-temporal dynamics.  (\aaa, \bbb, \ccc) Kymographs for a
fragment exhibiting organized dynamics with two consecutive backward flow waves
for each forward flow wave (reminiscent of a period doubling state).  (\ddd,
\eee, \fff) Kymographs for a fragment exhibiting disorganized dynamics.  }
\label{fig:kymo_other}
%
\end{figure}

Out of the 40 fragments in our study, 20 exhibited peristaltic behavior, 8 were
amphistaltic, and 2 alternated between peristaltic and amphistaltic.  In
addition, 5 fragments had organized spatio-temporal dynamics that did not match
either the peristaltic or amphistaltic patterns, and 5 more fragments had
disorganized dynamics.
Once a \phys fragment began migrating either by the peristaltic or by the
amphistaltic mode, the fragment would sustain the same mode for the duration of
the whole experiment, \ie $\gtrsim 30$ mins $\gtrsim 20$ cycles.  Thus, the
spatio-temporal dynamics of migrating \phys fragments appear to settle into
relatively robust oscillatory behaviors.
This observation generally agrees with Rodiek \etal \cite{rodiek2015patterns},
who measured the height oscillations of $\sim$ 1 mm--long \phys fragments while
they were migrating freely without being constrained by an agarose cap.  These
authors reported two spatio-temporal patterns in their measurements that
resemble the peristaltic and amphistaltic behaviors found in our experiments:
traveling waves that propagated at $\sim 5 \mu$m/s, and standing waves with
multiple spatial nodes separated by wavelength of $\sim 100\, \mu$m with period
of $10$ minutes.
We observed a few fragments that shifted spontaneously between the peristaltic
and the amphistaltic mode while migrating, as well as other organized
spatiotemporal patterns including 2-to-1 backward/forward flow waves (Figure
\ref{fig:kymo_other}\aaa,\bbb,\ccc) and disorganized patterns (Figure
\ref{fig:kymo_other}\ddd,\eee,\fff). 
This type of behavior is typical for systems with complex non-linear dynamics.
Consistently, previous experimental studies on \phys protoplasm droplets found
traveling waves, standing waves, and chaos in local droplet thickness
\cite{Takagi:2008,Takagi:2010}.  In addition, recent mathematical models that
include feedback between contraction, endoplasmic flow and calcium signaling in
simplified non-migratory geometries have predicted a number of dynamical
regimes depending on the level of mechano-chemical feedback and the rheological
properties of the endoplasm \cite{Radszuweit:14,Alonso:2016}.
%

\begin{figure}[tb!]
\begin{center}
\includegraphics[width=0.75\textwidth]{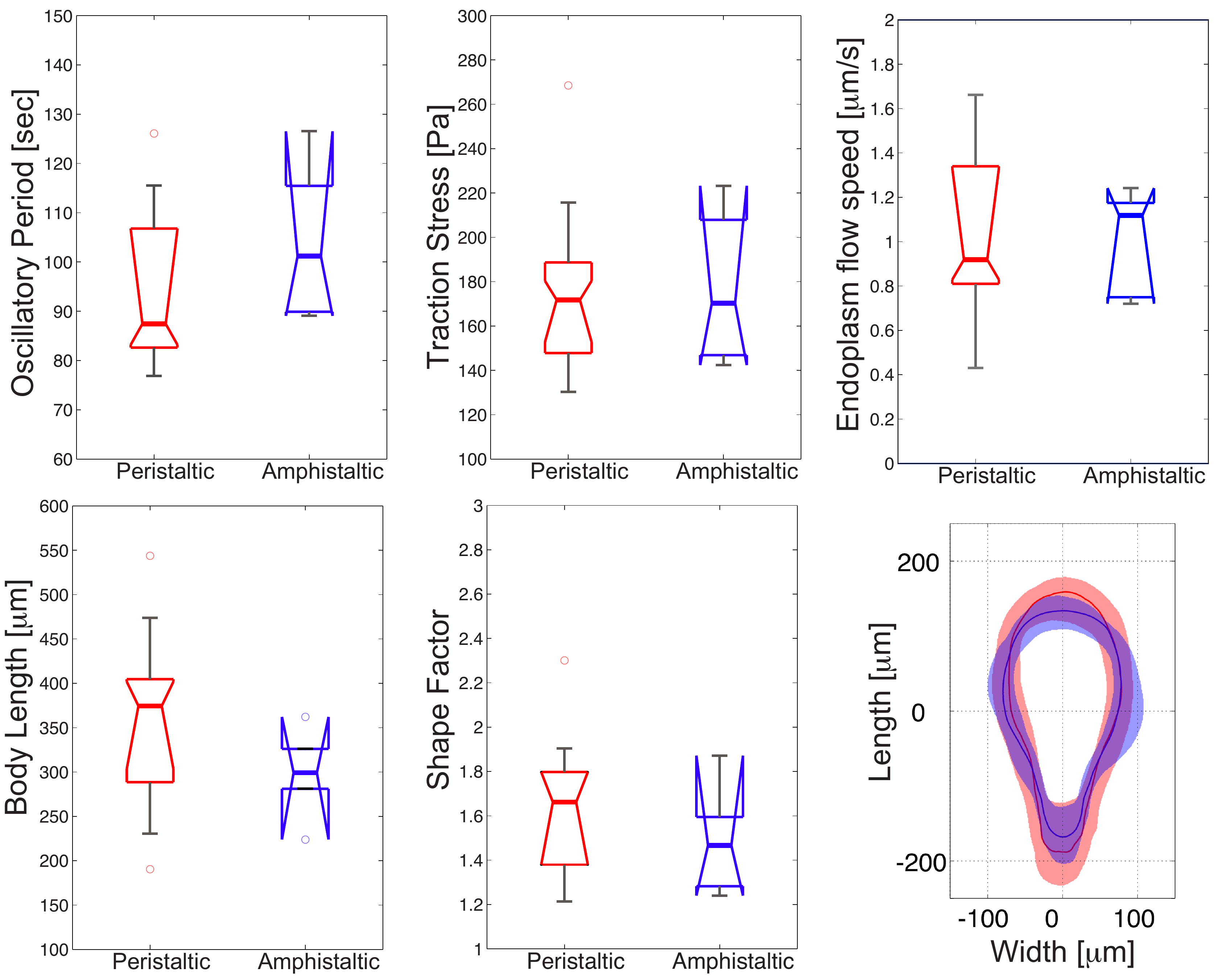}
\mylab{-0.7\textwidth}{0.575\textwidth}{{\footnotesize (\aaa)}}
\mylab{-0.455\textwidth}{0.575\textwidth}{{\footnotesize (\bbb)}}
\mylab{-0.215\textwidth}{0.575\textwidth}{{\footnotesize (\ccc)}}
\mylab{-0.73\textwidth}{0.27\textwidth}{{\footnotesize (\ddd)}}
\mylab{-0.485\textwidth}{0.27\textwidth}{{\footnotesize (\eee)}}
\mylab{-0.11\textwidth}{0.26\textwidth}{{\footnotesize (\fff)}}
\caption{ (\aaa)--(\eee) Box plots of motility parameters corresponding to
peristaltic ($N = 20$) and amphistaltic ($N = 8$) \phys fragments.
(\aaa) Average oscillation period.  (\bbb) Average
magnitude of the traction stresses.  (\ccc) Average endoplasmic flow speed.
(\ddd) Average fragment length.  (\eee) Shape factor.  (\fff) Average shape of
peristaltic (red line) and amphistaltic (blue line) types.  Shaded regions
contain 90\% of the statistical distribution of shapes for each fragment type.
}
\label{fig:boxplots}
\end{center}
\end{figure}

In an attempt to find differences in the properties of peristaltic and
amphistaltic \phys fragments that could explain their distinct dynamics, we
compared their oscillation period, average traction stress magnitudes and
average endoplasmic flow speeds.  However, we did not find any significant
difference in these parameters between the two types of fragments
(\Cref{fig:boxplots}\aaa-\ccc).  
Furthermore, both peristaltic and amphistaltic fragments adopted a similar
tadpole-like shape during migration (\Cref{fig:boxplots}\fff).  Rieu \etal
previously reported that \phys fragments with distinct dynamics of force
generation can be differentiated by the number of membrane invaginations
\cite{Rieu20150099}.  To quantify whether there were differences in the number
of membrane invaginations of \phys fragments undergoing peristaltic and
amphistaltic locomotion, we measured the shape factor $S_f = P^2(4\pi A)^{-1}$,
where $P$ is each fragment's perimeter and $A$ is its area.  This parameter is
unity for a perfect circle and increases as the number of lobes and
invaginations in the perimeter of the fragment increases.  No significant
difference was found between the two types of fragments
(\Cref{fig:boxplots}\eee).  
Ongoing measurements of the endoplasmic rheological properties
\cite{zhang2016feedback} should clarify if these properties play an important
role in establishing the spatio-temporal dynamical state of migrating \phys
fragments, as predicted by some mathematical models
\cite{Radszuweit:14,Alonso:2016}.
Despite these similarities, the average migration speeds of peristaltic and
amphistaltic fragments were significantly different, as shown in the next
section.

\subsection{The spatiotemporal dynamics of endoplasmic and ectoplasmic flows
affect the migration speed of \phys fragments}
\begin{figure}[tb!]
\begin{center}
\includegraphics[width=0.95\textwidth]{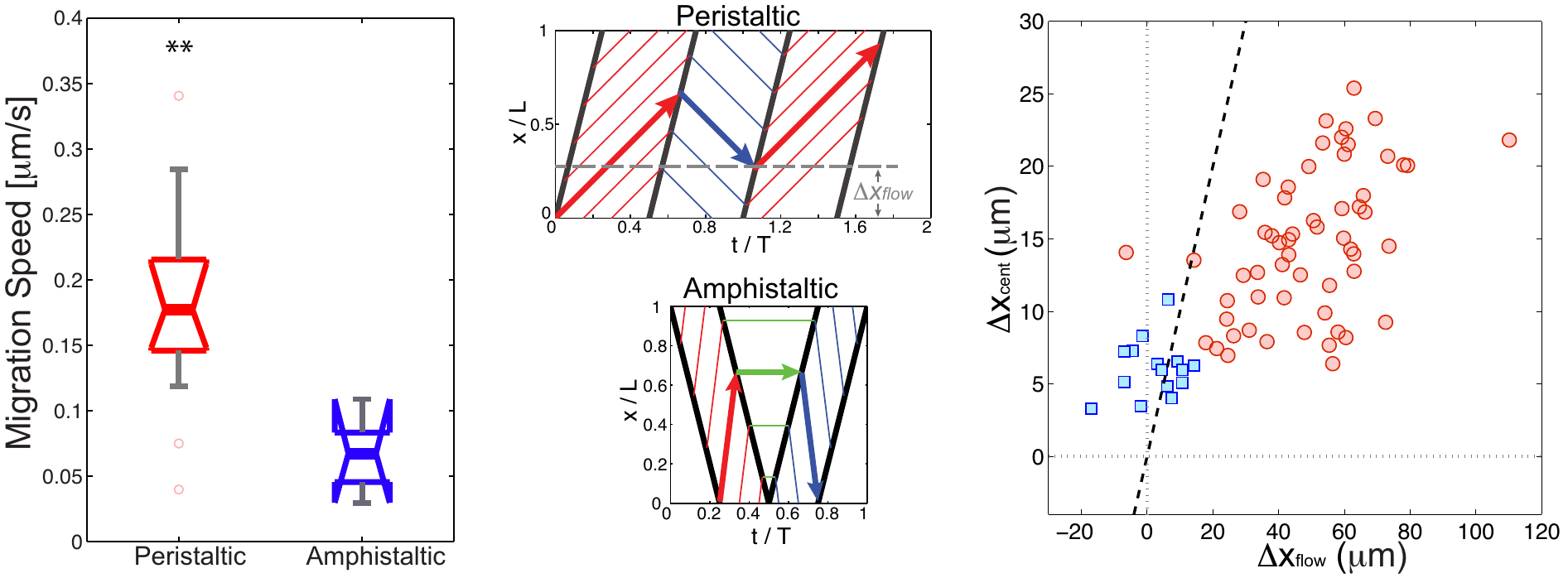}
\mylab{-0.89\textwidth}{0.32\textwidth}{{\footnotesize (\aaa)}}
\mylab{-0.675\textwidth}{0.32\textwidth}{{\footnotesize (\bbb)}}
\mylab{-0.32\textwidth}{0.32\textwidth}{{\footnotesize (\ccc)}}
\caption{ (\aaa) Box plot of average migration speeds in peristaltic ($N = 20$)
and amphistaltic ($N = 8$) \phys fragments.  Two asterisks denote statistically
significant differences between medians ($p < 0.01$).  
(\bbb) Simplified model schematic for the distance traveled by endoplasmic
fluid particles per oscillation cycle.  Top panel, peristaltic fragments;
bottom panel, amphistaltic fragments.
(\ccc) Scatter plot of  the distance $\Delta x_{cent}$ traveled by the centroid
of the \phys fragment per oscillation cycle vs.  the net distance $\Delta
x_{flow}$ traveled by an endoplasmic fluid particle.
\textcolor{red}{\solidcircle}, peristaltic fragments;
\textcolor{blue}{\solidsquar}, amphistaltic fragments.  The dashed line is
$\Delta x_{cent}=\Delta x_{flow}$.
}
\label{fig:scattaplo}
\end{center}
\end{figure}

We found that \phys fragments undergoing peristaltic migration were in average
$\sim 3$ times faster than those undergoing amphistaltic migration (Figure
\ref{fig:scattaplo}\aaa). 
This difference in locomotion speed is particularly remarkable considering that
both peristaltic and amphistaltic fragments have similar sizes and shapes, and
that their traction stresses and internal flow speeds have similar magnitudes
and oscillation periods (Figure \ref{fig:scattaplo}).
A possible explanation can be found by noting the different asymmetries in the
motion of endoplasmic fluid particles that arise from the different
spatio-temporal dynamics of flow waves in each locomotion mode.  
Figure \ref{fig:scattaplo} presents this idea by plotting spatio-temporal
particle trajectories in a simplified model in which the endoplasmic flow has
constant speed and waves propagate forward and backward with constant wave
speeds along the \phys fragment.
In peristaltic fragments, Matsumoto \etal \cite{Matsumoto:2008jf} noted that
fluid particles spend longer times traveling forward than backward, yielding
net forward displacement every cycle period ($\Delta x_{flow} > 0$, Figure
\ref{fig:scattaplo}\ccc).  Extending this argument to amphistaltic \phys
fragments predicts that fluid particles approximately return to their original
location at the end of each period (Figure \ref{fig:scattaplo}\ccc).
Consistent with this reasoning, a scatter plot of the net forward motion of the
centroid of a fragment ($\Delta x_{cent}$) vs. $\Delta x_{flow}$ clearly
segregates the amphistaltic and peristaltic locomotion modes (Figure
\ref{fig:scattaplo}\bbb).

It should be noted that the flow kinematics argument hold as long as the \phys
fragments do not experience shape changes over time scales longer than their
$\sim 100\;$s oscillation period.  This could explain Rodiek's \etal
\cite{rodiek2015patterns} observation that unconstrained \phys fragments that
sustain traveling waves in their height advance their front slower than
fragments that sustain multi-nodal standing waves, because their fragments
undergo substantial elongation and flattening during the duration of the
experiment.  In our experiments, this secular thickness variations are
constrained by agarose cap placed on top of the sample.

\begin{figure}[tb!]
\includegraphics[width=0.95\textwidth]{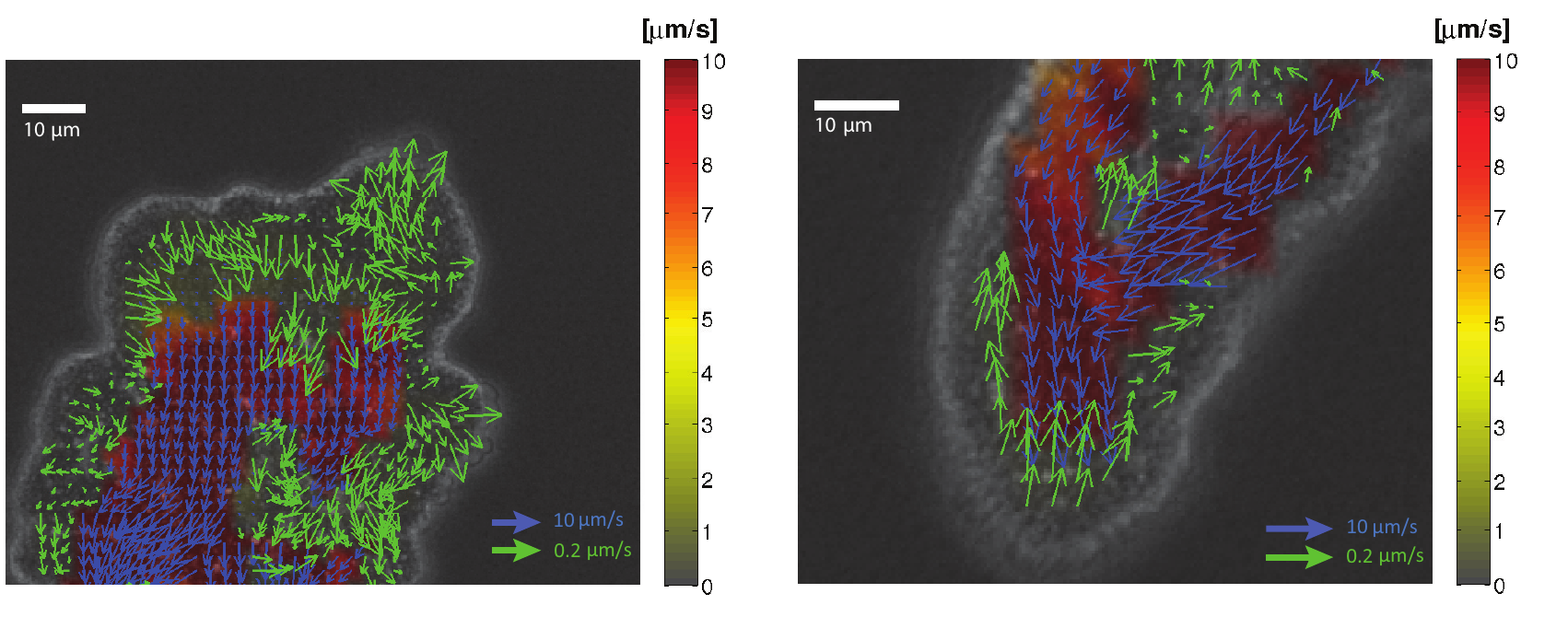}
\mylab{-0.95\textwidth}{0.36\textwidth}{{\footnotesize (\aaa) Fragment front}}
\mylab{-0.48\textwidth}{0.36\textwidth}{{\footnotesize (\bbb) Fragment rear}}
\vspace{-3ex}
\caption{Instantaneous snapshots showing velocity vectors for endoplasm (blue)
and ectoplasm (green) flows in a migrating \phys, superimposed on the bright
field image of the fragment.  The pseudo-color map indicates the magnitude of
velocity according to the colorbar in the right hand side of the
panel.
(\aaa) Frontal part of the fragment.  (\bbb) Rear part of the fragment.}
\label{fig:snapshot}
%
\end{figure}
While it provides a plausible explanation for the major differences in
migration speed found between peristaltic and amphistaltic fragments, the flow
kinematics hypothesis also poses a paradox because it predicts that
amphistaltic fragments should not be able to migrate.
Furthermore, we previously used mathematical modeling to show that the
asymmetry of endoplasmic flow velocity alone cannot determine the migration
speed of migrating \phys fragments \cite{Lewis20141359}.  The data in figure
\ref{fig:scattaplo}(\bbb), which shows that $\Delta x_{cent}$ is significantly
lower than $\Delta x_{flow}$, agrees with this idea.  
{\it Physarum} plasmodia are often conceptualized as being composed of a
two-phase fluid in which the sol and gel phases respectively represent the
endoplasm and the ectoplasm.  Therefore, it is reasonable to expect that the
dynamics of the ectoplasm may contribute to the net migration speed of the
plasmodium.

We still know little about the dynamics of the ectoplasm because its
motion is significantly slower and harder to measure than that of the
endoplasm.
%
%
%
In this study, we expanded the image processing algorithm for the
quantification of intracellular flow \cite{Lewis20141359}, in order to measure
the flow velocity of the ectoplasm in addition to that of the endoplasm (see \S
\ref{sec:methods} and Figure \ref{fig:snapshot}).
%
%

%

\begin{figure}[tb!]
\includegraphics[width=0.95\textwidth]{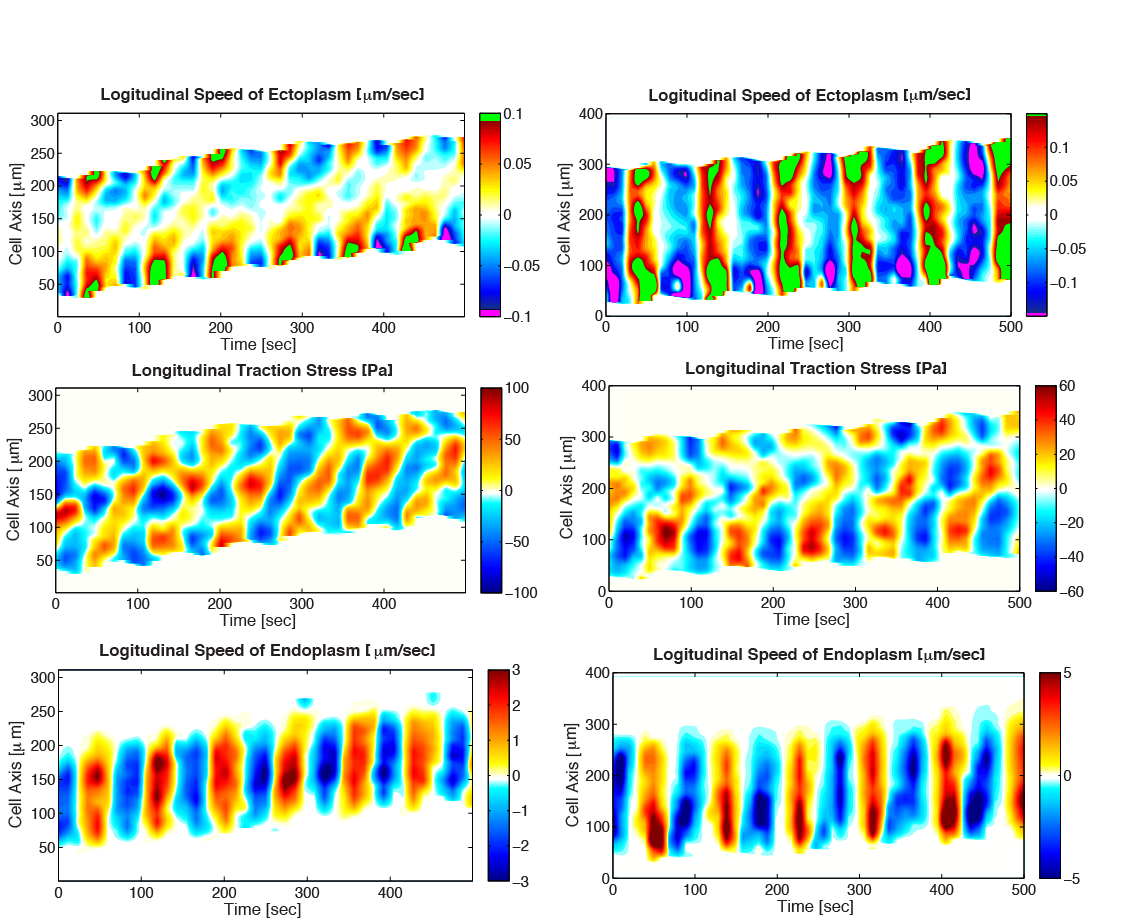}
\mylab{-0.9\textwidth}{0.685\textwidth}{{\footnotesize (\aaa) }}
\mylab{-0.91\textwidth}{0.445\textwidth}{{\footnotesize (\bbb) }}
\mylab{-0.915\textwidth}{0.195\textwidth}{{\footnotesize (\ccc) }}
\mylab{-0.445\textwidth}{0.685\textwidth}{{\footnotesize (\ddd) }}
\mylab{-0.455\textwidth}{0.445\textwidth}{{\footnotesize (\eee) }}
\mylab{-0.455\textwidth}{0.195\textwidth}{{\footnotesize (\fff) }}
\vspace{-3ex}
\caption{Kymographs of longitudinal ectoplasm flow velocity (\aaa, \ddd),
longitudinal traction stress (\bbb, \eee), and longitudinal endoplasm flow
velocity (\ccc, \fff) for a peristaltic (\aaa--\ccc) and an amphistaltic
(\ddd--\fff) \phys fragment. 
In panels (\aaa, \ddd), we have added bright green and purple at the floor and
ceiling of the colormaps to emphasize asymmetries in the velocity data.
}
\label{fig:kymo2}
%
\end{figure}

Ectoplasm velocity is organized spatio-temporally in the form of traveling
waves and standing waves in peristaltic and amphistaltic fragments respectively
(Figure \ref{fig:kymo2}\aaa,\ddd), consistent with the dynamics of the traction
stresses generated by the fragments (\Cref{fig:kymo2}\bbb,\eee) and their endoplasm
flow velocity (\Cref{fig:kymo2}\ccc,\fff).  However, in both types of fragments the
ectoplasm velocity is asymmetric reaching substantially higher values during
forward motion than during backward motion, particularly in the rear.  These
results suggest that the dynamics of the ectoplasm also contribute to the net
motion of \phys fragments.

\subsection{Dynamics of substratum adhesion experience smooth slip-stick
transitions}
%

%
\begin{figure}[tb!]
\includegraphics[width=0.95\textwidth]{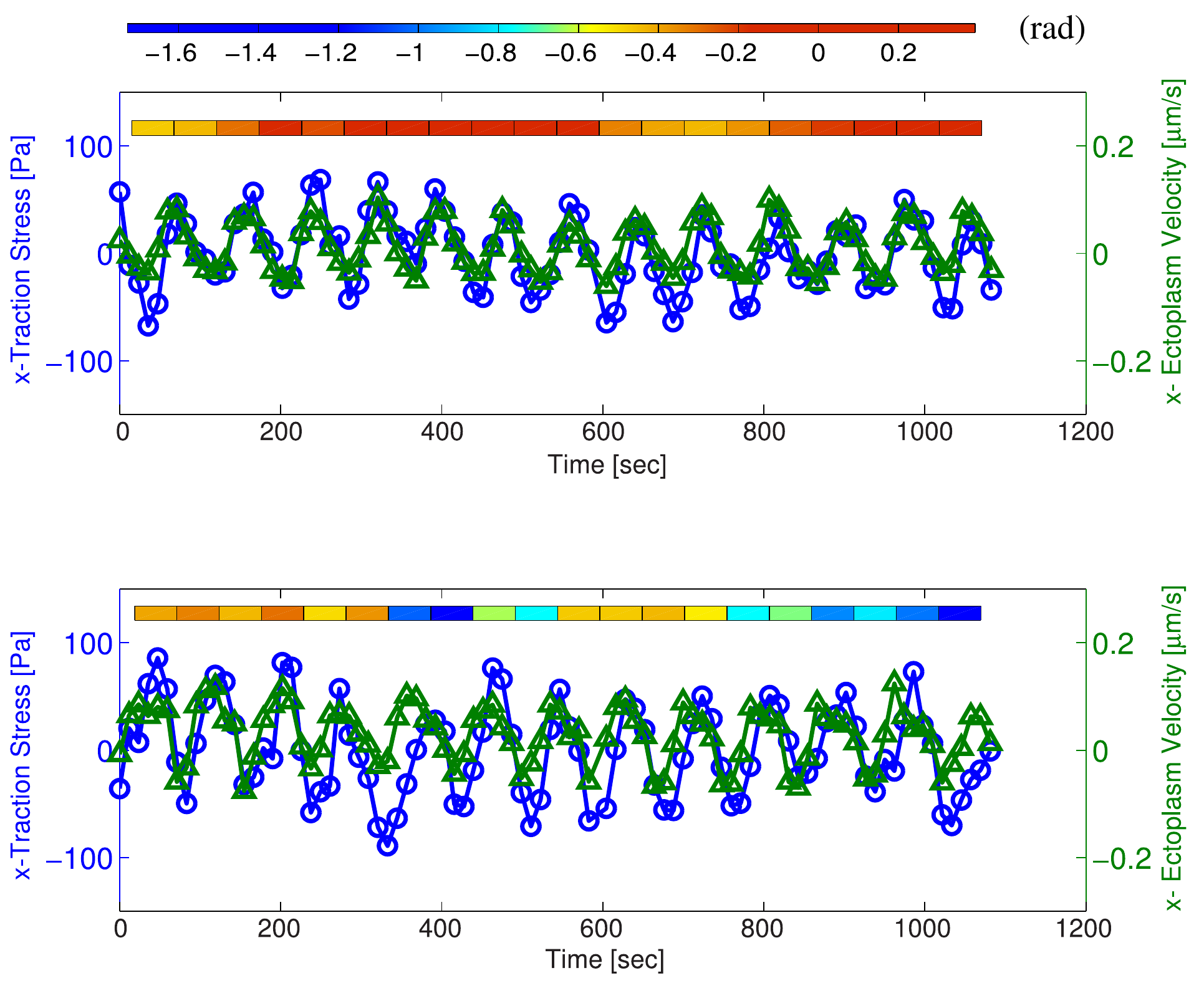}
\mylab{-0.87\textwidth}{0.66\textwidth}{{\footnotesize (\aaa) Front}}
\mylab{-0.875\textwidth}{0.32\textwidth}{{\footnotesize (\bbb) Rear}}
\vspace{-2ex}
\caption{Time histories of longitudinal ectoplasm velocity
(\textcolor{verde}{\linetri}) and longitudinal traction stresses
(\textcolor{blue}{\linecir}) at two specific locations in the front (panel
\aaa) and the back (panel \bbb) of the peristaltic \phys fragment shown in
Figure \ref{fig:kymo2}.
The tiled bars at the top of the plots represent the
time-dependent phase differences (in radians) between the ectoplasm velocity
and the traction stresses.  Blue and orange tiles represent phase differences
near $-\pi/2$ (cell and substrate stick) and zero (cell and substrate slip)
respectively, as indicated by the color scale at the top of the figure.} 
\label{fig:peri_all}
\end{figure}

Inspection of Figure \ref{fig:kymo2} suggests that traction stresses generation
is closely related to the motion of the ectoplasm over the substratum, which is
sound considering that the ectoplasm forms a cortical layer directly in contact
with the plasmodial membrane.  We explored this relationship in more detail in
order to gain insight about the regulation of substratum adhesion in migrating
\phys fragments, which is not well understood since integrin-like adhesion
proteins have not been identified yet for this organism.

We plotted the time evolution of the ectoplasm speed together
with that of the traction stresses at the front of a fragment
(\Cref{fig:peri_all}\aaa), and with the phase difference between these two
variables.
The time lag between the two signals was calculated by maximizing their
cross-correlation over interrogation windows of 95 seconds and 50\% overlap.
The instantaneous phase difference was then obtained as the ratio of time lag
and the averaged oscillation period (83 seconds).
This analysis revealed that the traction stresses and ectoplasm velocity 
oscillate in phase for the most part.  This result suggest that adhesion in
this region follows the viscous-like regime $\tau = \xi v$ where $\tau$ is the
adhesion stress, $v$ the ectoplasm velocity and $\xi$ is a friction factor.  
This result is in agreement with theoretical studies describing biological
friction as the consequence of the thermally driven formation and rupture of
molecular bonds \cite{Srinivasan:2009cg}.

In contrast to our observations at the front of the fragments, the time
evolutions of ectoplasm speed and traction stresses have a complex relationship
at the rear, alternating intervals at which they oscillate in phase with
intervals in which the  ectoplasm velocity precedes the traction stresses by
approximately $1/4$ of an oscillation period.  Since a time integration of
endoplasm velocity (\ie endoplasm displacement) would generate the same phase
difference of $1/4$ period, we interpret this result as indicative of the
occurrence of stick-slip transitions at the rear of \phys fragments.
Slip-stick transitions occur when $\xi$ has a non-monotonic dependence with the
ectoplasm velocity, which is a common behavior in biological friction
\cite{Srinivasan:2009cg}.  This type of transitions have been linked to the
dramatic shape oscillations experienced by cells such as keratocytes when
crawling on flat substrata \cite{Barnhart:2010}, and it has been proposed that
the frequency of these oscillations correlates with the speed of cell crawling
\cite{DelAlamo:2007jm,meili:2010,Barnhart:2010,bastounis:2011scar}.
In the present experiments, we did not observe sharp changes in traction
stress, ectoplasm speed or fragment length occurring at the stick-slip
transitions, suggesting that these transitions are mild in migrating \phys
fragments.  We analyzed kymographs of the phase difference between the time
evolutions of traction stress and ectoplasm speed (not shown).  While these
data were somewhat noisy, they suggested that periodic stick-slip transitions
propagate from the rear to the front of peristaltic fragments, while a standing
stick-slip transition seems to form near the front of amphistaltic fragments.
The dynamics of these transitions could provide a mechanism for \phys to
regulate the strength of their substrate adhesion in a way that supports
asymmetry in the motion of the ectoplasm.  However, additional experiments and
further analysis are needed to confirm these ideas.

\subsection{Dynamics of free intracellular calcium}
\label{sec:ca2plus}
\begin{figure}[tb!]
\includegraphics[width=0.95\textwidth]{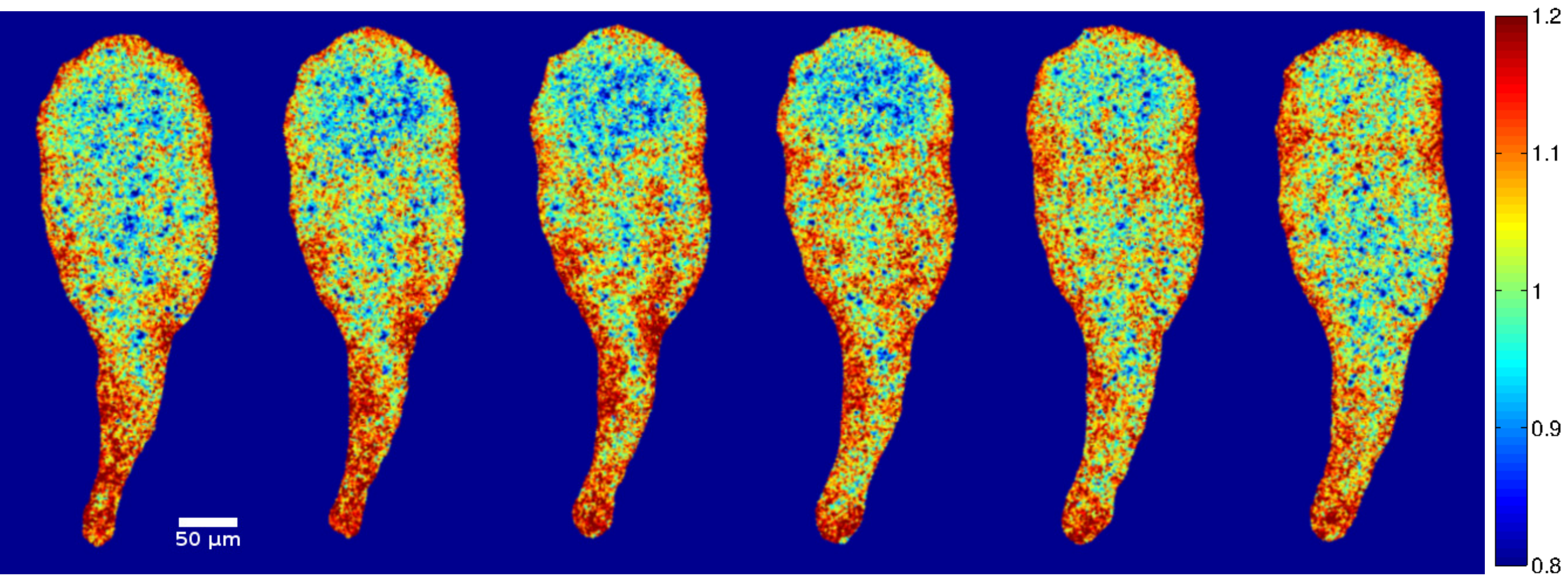}
\vspace{-2ex}
\caption{Time sequence of ratiometric measurememt of [Ca$^{2+}$]i during the
locomotion of a typical peristaltic cell showing a Ca$^{2+}$ wave propagating
forward.}
\label{fig:snapshots4}
\end{figure}

The transport of calcium ions in \phys fragments occurs in a complex regime
that likely couples convection, diffusion and a time-dependent geometry caused
by fluid-structure interactions at the fragment's lengthscale.  Using reported
values of cytoplasmic Ca$^{2+}$ diffusivity, $D= 5.3 \times 10^{-10} m^2/s$
\cite{Donahue:1987}, and our measurement of intracellular flow speed $v \sim 5
\mu m/s$, we estimate that characteristic timescales for Ca$^{2+}$ diffusion
and convection over a cell length ($l \sim 100 \mu m$) are the same, $t_D = t_C
= 20 s$.  Furthermore, this transport timescale is similar to the period of
cellular shape changes ($T \approx 100 s$) observed in our experiments.  

\begin{figure}[tb!]
\includegraphics[width=0.95\textwidth]{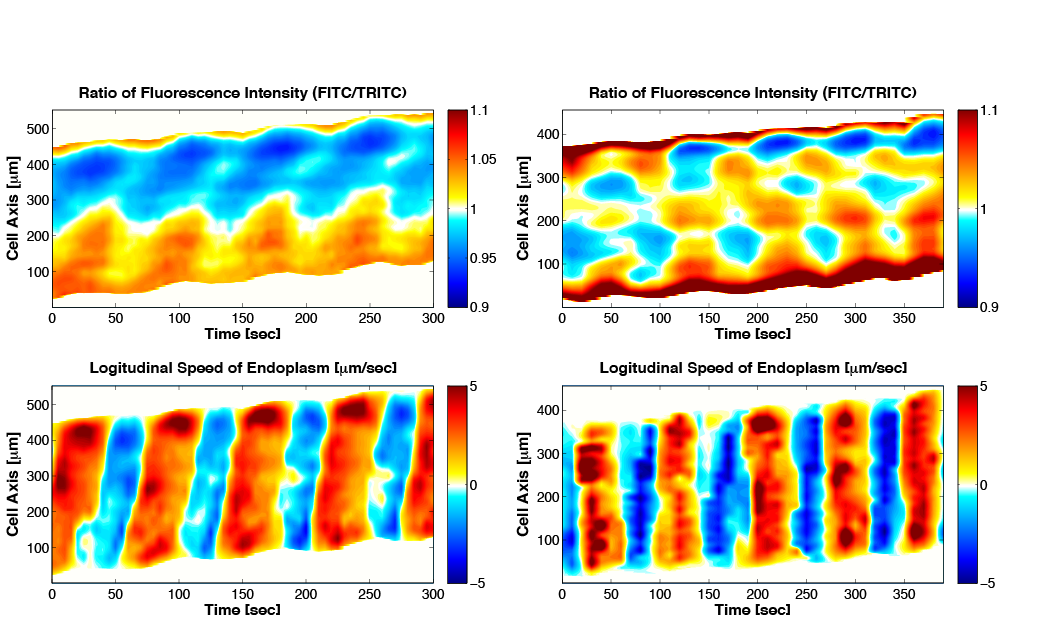}
\mylab{-0.94\textwidth}{0.5\textwidth}{{\footnotesize (\aaa)}}
\mylab{-0.94\textwidth}{0.24\textwidth}{{\footnotesize (\bbb)}}
\mylab{-0.48\textwidth}{0.5\textwidth}{{\footnotesize (\ccc)}}
\mylab{-0.48\textwidth}{0.24\textwidth}{{\footnotesize (\ddd)}} \caption{(\aaa)
Kymograph of ratiometric measurement of [Ca$^{2+}$]i in a typical peristaltic
fragment.  (\bbb) Kymograph of instantaneous longitudinal velocities of
endoplasmic flow of the same peristaltic fragment in \Cref{fig:kymo4}(\aaa).  (\ccc)
Kymograph of ratiometric measurement of [Ca$^{2+}$]i in a typical amphistaltic
fragment.  (\ddd) Kymograph of instantaneous longitudinal velocities of
endoplasmic flow of the same amphistaltic fragment in \Cref{fig:kymo4}(]\ccc).}
\label{fig:kymo4}
\end{figure}

In order to study the relation between endoplasmic flow and the distribution of
free intracellular calcium, we performed ratiometric measurements of free ion
concentration, [Ca$^{2+}$]i, jointly with intracellular flow.
\Cref{fig:snapshots4} shows a time sequence of [Ca$^{2+}$]i throughout one
oscillation cycle of a typical peristaltic cell.  Although this type of
measurement is inherently noisy, it is possible to discern waves of high
calcium concentration propagating from the rear to the front of the \phys
fragment.  The spatio-temporal dynamics of these waves are clearly observed
when [Ca$^{2+}$]i is represented in kymographic form. 
\Cref{fig:kymo4} shows kymographs of [Ca$^{2+}$]i and intracellular flow that are
representative of the peristaltic and amphistaltic migration modes.
%
%

Both for the peristaltic and amphistaltic modes, the spatio-temporal patterns
of calcium concentration are consistent with the dynamics of the traction
stress, endoplasmic flow and ectoplasmic motion.  
In \phys fragments undergoing peristaltic migration, we found waves of
[Ca$^{2+}$]i that traveled from the rear to the front of the fragment
(\Cref{fig:kymo4}\aaa), whereas patterns of [Ca$^{2+}$]i standing waves were
observed for \phys fragments undergoing amphistaltic locomotion
(\Cref{fig:kymo4}\ccc).  
The phase speed of the traveling waves was found to agree well with the
measured endoplasmic flow velocity, $v_0 \approx 5\, \mu m /s$, suggesting that
endoplasmic flows may be important in sustaining the dynamics of [Ca$^{2+}$]i
transport.

\begin{figure}[tb!]
%
\vspace{1ex}
\includegraphics[width=0.95\textwidth]{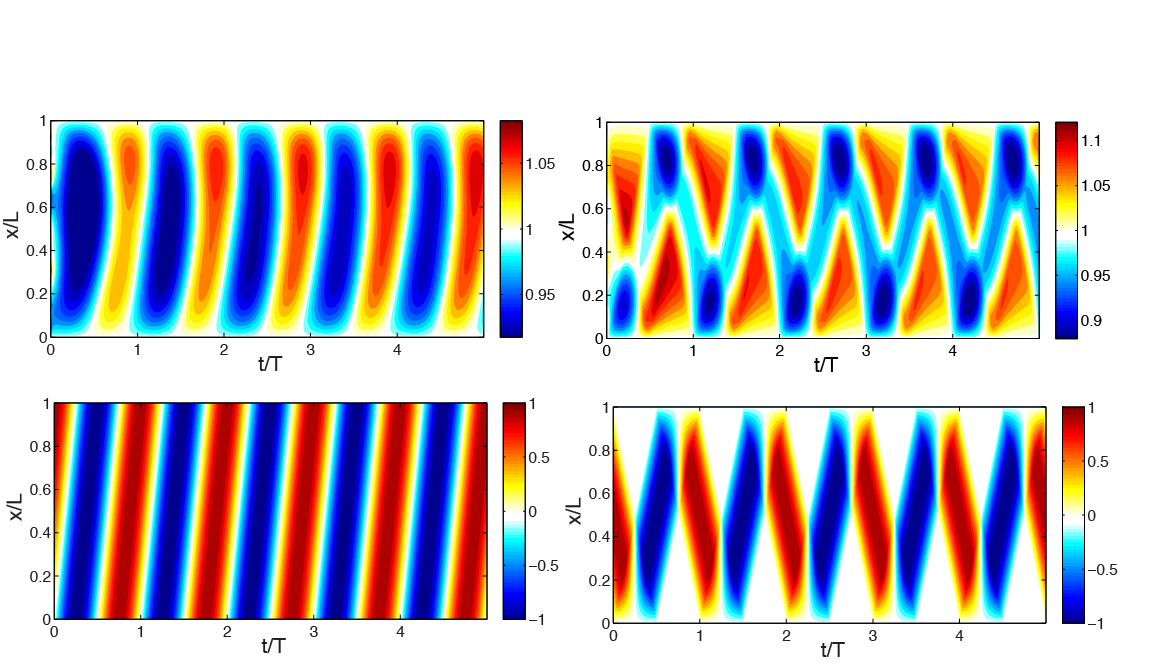}
\mylab{-0.91\textwidth}{0.48\textwidth}{{\footnotesize (\aaa) Passive Scalar}}
\mylab{-0.91\textwidth}{0.23\textwidth}{{\footnotesize (\bbb) Endoplasm Flow}}
\mylab{-0.46\textwidth}{0.48\textwidth}{{\footnotesize (\ccc) Passive Scalar}}
\mylab{-0.46\textwidth}{0.23\textwidth}{{\footnotesize (\ddd) Endoplasm Flow}}
\caption{(a) Kymograph of concentration of passive scalar in a mimic
peristaltic fragment.  (b) Kymograph of longitudinal velocities of endoplasmic
flow representative of a peristaltic fragment.  (c) Kymograph of concentration
of passive scalar in a mimic amphistaltic fragment.  (d) Kymograph of
longitudinal velocities of endoplasmic flow of endoplasmic flow representative
of an amphistaltic fragment.}
\label{fig:kymo5}
\end{figure}

To test this hypothesis, we considered a simple 1-D model for the transport of
a passive scalar in non-dimensional form,
\begin{equation}
\mbox{St}\, \partial_t c + v_{endo} \partial_x c = \mbox{Pe}^{-1} \partial_{xx}
c
\end{equation}
where $v_{endo}$ is a prescribed velocity normalized with $v_0$, the spatial
variable $x\, \epsilon [0,1]$ is normalized with the fragment length $L$, and
the time variable $t$ is normalized with the period of the flow oscillations
$T$.  The two non-dimensional parameters in this equation are the Strouhal
number $\mbox{St} = L / (v_0 T)$ and the P\'eclet number $\mbox{Pe}= L v_0 /
D$, both of which have values of order unity in migrating \phys fragments
according to our experimental measurements and \cite{Donahue:1987}.
The solution to this transport equation exhibits traveling waves of passive
scalar when $v_{endo}$ is set to mimic our experimental measurements for
peristaltic fragments, \eg $v_{endo} = \sin[2\pi(x - \zeta t)]$ where the
non-dimensional phase velocity $\zeta \approx 5$ (Figure
\ref{fig:kymo5}\aaa--\bbb).  Likewise, this simple model generates standing
waves of passive scalar when the endoplasmic velocity is set to mimic our
measurements for amphistaltic fragments (Figure \ref{fig:kymo5}\ccc--\ddd).
Qualitatively, these results are robust with respect to changes in the
parameter values, and in the boundary conditions (e.g. Neumann vs. Dirichlet)
and initial conditions.  For instance, the passive scalar in Figure
\ref{fig:kymo5} evolves from random initial conditions into temporally periodic
pattern in about one oscillation cycle.
%

%

%
It is evident that a flow-transport-only model for the dynamics of
intracellular calcium does not capture many of the quantitative features
observed in our measurements of Figure \ref{fig:kymo4}.
It is also evident that such a simplistic model neglects potentially relevant
phenomena such as chemical kinetics of phosphorylation/dephosphorylation of the
myosin light chain, Ca$^{2+}$ influx through various calcium channels on plasma
membrane, Ca$^{2+}$ 
\textcolor{blue}{release}
from the sarcoplasmic reticulum and endoplasmic reticulum through IP$_{3}$
channels, etc
\cite{kamm1985function,karaki1997calcium,endo1977calcium,watras1991bell,
clapham2007calcium,rizzuto2006microdomains}.
Nevertheless, the ability of such a simple model to generate traveling waves
and standing waves of a passive scalar highlight the importance of endoplasmic
flow in the self-organization of dynamical patterns in migrating \phys
fragments.

The molecular regulation of acto-myosin contractility by calcium should be the
same for \phys fragments following the peristaltic and amphistaltic migration
modes, given that we prepared all the fragments using the same protocol and the
emergence of these modes was spontaneous.  Thus, we hypothesized that the phase
coordination between the dynamics of calcium and contractility waves would be
the same for both migratory modes.
%
\begin{figure}[tb!]
%
\vspace{2ex}
\includegraphics[width=0.99\textwidth]{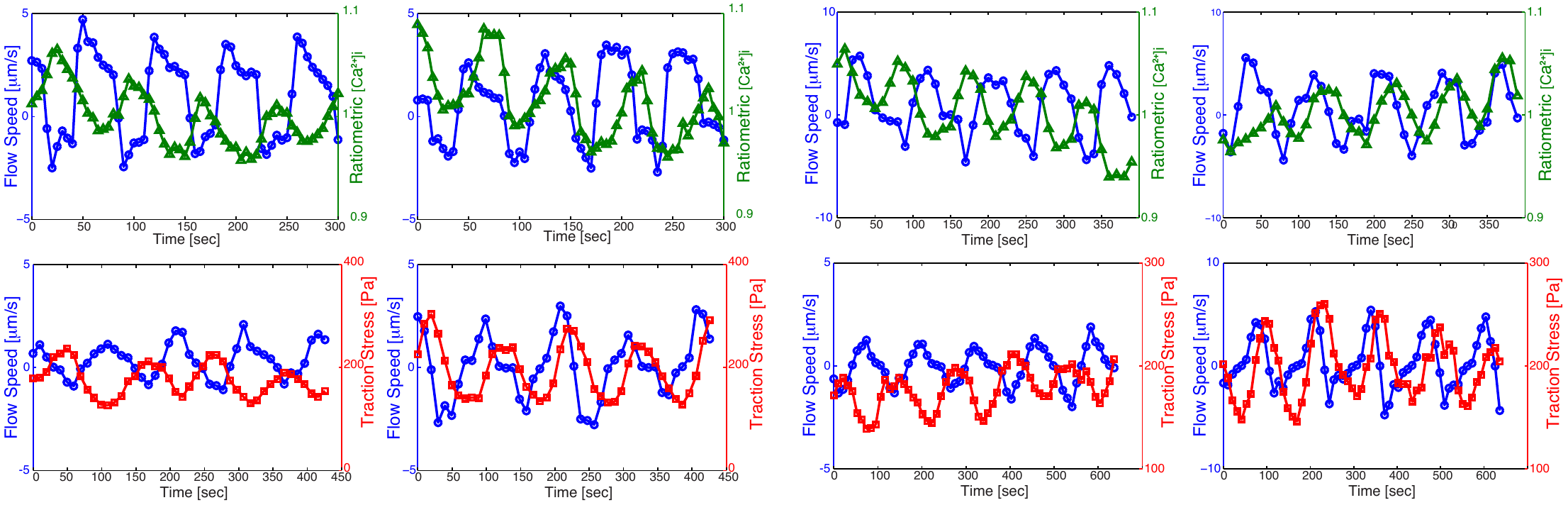}
\mylab{-0.98\textwidth}{0.32\textwidth}{{\footnotesize (\aaa)}}
\mylab{0.26\textwidth}{0.355\textwidth}{{\footnotesize (\ccc)}}
\mylab{0.52\textwidth}{0.355\textwidth}{{\footnotesize (\eee)}}
\mylab{0.76\textwidth}{0.355\textwidth}{{\footnotesize (\ggg)}}
\mylab{0.0\textwidth}{0.17\textwidth}{{\footnotesize (\bbb)}}
\mylab{0.235\textwidth}{0.17\textwidth}{{\footnotesize (\ddd)}}
\mylab{0.49\textwidth}{0.17\textwidth}{{\footnotesize (\fff)}}
\mylab{0.725\textwidth}{0.17\textwidth}{{\footnotesize (\hhh)}}
\vspace{-3ex}
\caption{Top row (\aaa, \ccc, \eee, \ggg): Time histories of endoplasmic flow
velocity (\textcolor{blue}{\linecir}) and ratiometric measurement of
[Ca$^{2+}$]i (\textcolor{verde}{\linetri}), averaged along the width of a
peristaltic fragment (panels \aaa\; and \ccc) and an amphistaltic fragment
(panels \eee\; and \ggg).
Bottom row (\bbb, \ddd, \fff, \hhh): Time histories of endoplasmic flow
velocity (\textcolor{blue}{\linecir}) and peripheral traction stress
(\textcolor{red}{\linesquar}), averaged along the width of a peristaltic
fragment (panels \bbb\; and \ddd) and an amphistaltic fragment (panels \fff\;
and \hhh).
Panels (\aaa, \bbb, \eee, \fff): Fragment front.  Panels (\ccc, \ddd, \ggg,
\hhh): Fragment rear.}
\label{ca_peri_all}
\end{figure}
While it was not possible to measure [Ca$^{2+}$]i and traction stresses
simultaneously in our experiments, we measured both [Ca$^{2+}$]i jointly with
endoplasmic flow, and traction stresses jointly with endoplasmic flow.  The
flow data were then used as reference to temporally align the oscillations of
[Ca$^{2+}$]i and traction stress for \phys fragments following the same
migration mode.
We plotted time profiles at specific locations at the front and rear of each
fragment (\Cref{ca_peri_all}), and juxtaposed the time evolution of
endoplasmic flow velocity to that of [Ca$^{2+}$]i or traction stresses
(\Cref{ca_peri_all}).

For the peristaltic migration mode, the time evolutions of endoplasmic flow and
[Ca$^{2+}$]i were found to have opposite phases at the front of the \phys
fragment (\Cref{ca_peri_all}\aaa).  The time evolutions of endoplasmic flow and
traction stress also had opposite phases at the fragment front
(\Cref{ca_peri_all}\bbb), implying that the oscillations in [Ca$^{2+}$]i and
traction stress were in phase.  The same relationship between [Ca$^{2+}$]i,
endoplasmic flow and the traction stresses can be deduced from the time
profiles of these variables recorded at the rear of peristaltic \phys fragments
(\Cref{ca_peri_all}\ccc, \ddd).
In amphistaltic fragments (\Cref{ca_peri_all}\eee--\hhh), the phase
coordination between the waves of [Ca$^{2+}$]i, flow speed and traction stress
was the same as in peristaltic fragments.
Our finding that the calcium concentration is in phase with traction stresses
agrees with previous observations
\cite{yoshiyama2010calcium,zhang2012calcium,nakamura1999calcium}.  Yoshiyama
\etal \cite{yoshiyama2010calcium} interpreted this result as an indication that
calcium inhibits acto-myosin contractility in {\it Physarum} because the
maximum calcium concentration coincides with the onset of relaxation.  However,
the kinetics of the involved biochemical reactions could make this response
more complicated \cite{smith1992model}.

\section{Conclusion}

The multi-nucleated slime mold \phys {\it polychephalum} can be used to
generate amoeboid-like motile cells by excision of $\sim 100\; \mu$m-long
fragments from the parent mold.  These fragments are formed by a cortical
gel-like ectoplasm that surrounds a sol-like endoplasm.  Periodic contractions
of the ectoplasm drive shuttle flows in the endoplasm, which transport the
nutrients and calcium ions necessary for contraction.  The feedback among these
processes can lead to rich spatio-temporal dynamics that significantly affect
the migration behavior of \phys fragments.  However, our understanding of these
dynamics is limited by a lack of direct measurements of quantitative variables.
This study provides detailed concurrent measurements of the spatio-temporal
distribution of endoplasmic and ectoplasmic flow, contractile forces and
[Ca$^{2+}$]i in migrating fragments of \phys plasmodia.  To the best of our
knowledge, this is the first experimental quantification of mechano-chemical
dynamics in a model organism of flow-driven amoeboid migration. 

The spatio-temporal patterns found in the measured quantities suggests that the
mechano-chemical dynamics of these fragments can lead to a variety of both
disorganized and organized states.  We focused our attention on two
particularly stable organized states associated with periodic oscillations in
flow, contractile forces and [Ca$^{2+}$]i.  
In the most stable (\ie frequently observed) state, the mechano-chemical
dynamics of the fragment are organized in the form of traveling waves that
propagate from the rear to the front of the fragment, in good agreement with
the {\it peristaltic} behavior studied in previous works
\cite{Matsumoto:2008jf,Lewis20141359}.  
We also investigated a second stable dynamical state that we termed {\it
amphistaltic} because it consists of alternate anti-phase contractions and
relaxations of the fragment's front and back (from $\alpha \mu \phi \i$ in
Greek meaning ``on both sides'').  These anti-phase contractions are associated
with standing waves of traction stresses and [Ca$^{2+}$]i, but they lead to
traveling waves of endoplasmic flow with alternating propagation directions;
waves of forward flow propagate backward and viceversa, leading to clear
$V$-shape patterns in spatio-temporal flow kymographs.

Our data suggest that the transport of calcium ions by the endoplasmic flows
observed in \phys fragments may be fundamental to coordinate the
spatio-temporal patterns of traction stresses that drive their locomotion.
Specifically, we showed that the forward traveling waves of endoplasmic flow
found in peristaltic fragments can sustain traveling waves of calcium
concentration, consistent with our experimental measurements of [Ca$^{2+}$]i.
In a similar fashion, the flow waves of alternating propagation direction
observed in amphistaltic fragments can sustain standing waves of [Ca$^{2+}$]i,
also consistent with our experimental measurements.  Furthermore, we showed
that the patterns of concentration of calcium ions evolve in space and time
with the same phase as those of the traction stresses.

Apart from the organization of their mechano-chemical dynamics, we did not
observe significant differences between the properties of peristaltic and
amphistaltic \phys fragments.  Both types of fragments were found to have
similar sizes and shapes, and their traction stresses and internal flow speeds
were found to have similar magnitudes and oscillation periods.  Nevertheless,
the average migration speed of peristaltic fragments was measured to be $3$
times higher than the migration speed of amphistaltic fragments.
We argued that this difference could be caused in part by the spatio-temporal
dynamics of the endoplasmic flows in the two types of fragments.  In
peristaltic fragments both positive and negative endoplasmic velocity waves
propagate forward, which allows for positive net endoplasmic motion every
oscillation cycle \cite{Matsumoto:2008jf}.  On the other hand, in amphistaltic
fragments positive and negative velocity waves propagate in opposite
directions, which leads to zero net endoplasm motion.  

Albeit slowly, amphistaltic fragments undergo peristent directional migration
over long periods of time. Thus, it is evident that analyzing the symmetry of
endoplasm flow is not sufficient to capture the migratory behavior of \phys
fragments.  In contrast to the endoplasm, we observed that the ectoplasm of
both peristaltic and amphistaltic fragments flows faster forward than it does
backward, leading to a significant amount of net motion per cycle.  In a
previous study \cite{Lewis20141359}, we used numerical modeling to illustrate
that this type of asymmetry in ectoplasm motion would require tight
coordination between the generation of contractile forces and the adhesion of
the ectoplasm to the substratum.  
We explored this coordination by comparing the time evolutions of the traction
stresses generated by the \phys fragments and of the motion of their ectoplasm.
Our experimental results suggest that the spatio-temporal coordination between
these two quantities may be realized by means of stick-slip transitions.  
We occasionally observed stationary “hotspots” in the measured traction
stresses (see \eg Figure \ref{fig:snapshots0}\aaa), which might be associated
with the stick-slip transitions observed at the fragment tail.
These stationary adhesions are however more common in other types of amoeboid
cells such as \emph{Amoeba proteus} or \emph{Dyctiostelium discoideum}.  In
both cell types, the adhesion sites remain stationary as the cells migrate over
them, leaving a clear signature in the kymographs that consists of horizontal
bands (see Supplementary Information and refs.
\cite{Bastounis:2014gg,AlvarezGonzalez:2015km}) This behavior significantly
contrasts with the dynamics of the traction stresses observed in migrating
\emph{Physarum} fragments.
Finally, our data provide preliminary evidence that these transitions might be
organized in the form of traveling or standing waves, consistent with the
dynamics of endoplasmic flow, contractility and calcium distribution.  Further
experiments and analyses are needed to confirm and expand this mechanistic
model.  \\[3ex]

This work was supported by the National Science Foundation through grant
CBET-1055697 and by the National Institute of Health via grants R01-GM084227
and R01-HL128630.

\newpage

\end{document}